\documentclass[aps,rmp,preprint,amsmath,amssymb,graphicx,longbibliography,floatfix,showpacs]{revtex4-1}
\begin{document}

\title{Introduction to the Special Issue on the Statistical Mechanics of Climate}

\author{Valerio Lucarini}

\affiliation{Department of Mathematics and Statistics, University of Reading, Reading UK}
\affiliation{Centre for the Mathematics of Planet Earth, University of Reading, Reading, UK}
\affiliation{CEN, University of Hamburg, Hamburg, Germany}
\email[Email: ]{v.lucarini@reading.ac.uk}
\date{\today}

\begin{abstract}
We introduce the special issue on the \textit{Statistical Mechanics of Climate} published on the Journal of Statistical Physics by presenting an informal discussion of some theoretical aspects of  climate dynamics that make it a topic of  great interest for mathematicians and theoretical physicists. In particular, we briefly discuss its nonequilibrium and multiscale properties, the relationship between natural climate variability and climate change,  the different regimes of climate response to perturbations, and critical transitions.

\end{abstract}

\maketitle 

\maketitle

The activities in the research area at the intersection between mathematics, theoretical physics, and Earth system science have received a very powerful impulse following the Mathematics of Planet Earth 2013 (MPE2013) international scientific programme (see \texttt{http://mpe.dimacs.rutgers.edu}), which has paved the way for many scientific initiatives and funding opportunities, and, more generally, for bringing to the spotlight  a vast range of interdisciplinary research activities of great relevance in terms of science \textit{per se} as well as of socio-environmental challenges they can contribute to.  Obviously, there is a long history of two-way interactions between Earth system sciences, on the one hand, and mathematics and theoretical physics, on the other hand – see the examples of chaos theory, (geophysical) fluid dynamics, fractals, extreme value theory, stochastic dynamical systems, data assimilation, just to name a few. Additionally, since the very start of the computer age, modelling exercises dedicated to the simulation of the weather, and later of the ocean and of the climate as a whole have consistently been some of the heaviest users of high-performance computing.
Finally, the current revolution in data science is finding very important applications in Earth system science as well
as receiving many challenging inputs from it \cite{Hosni2018,Buchanan2019,Faranda2019}. 

The goal of this special issue is to contribute to such interdisciplinary challenges by hosting scientific contributions that, on the one hand, try to move forward – through theory, numerical simulations, and analysis of data – the understanding of the climate system through the lens of statistical mechanics and, on the other hand, develop ideas of relevance for mathematics and physics taking inspiration from problems emerging in climate science. 

Using mostly a plain language, we introduce here some  theoretical aspects of  climate dynamics that make it a topic  of  great interest for mathematicians and theoretical physicists, and try to convince the reader of the existence of a great potential for important results both at fundamental level and in terms of usable tools for studying specific problems associated with the understanding of the dynamics of the climate system. A  more detailed exposition of the topics presented here (and of much more) can be found in a recent review paper \cite{GhilLucarini2020}; see also \citet{Lucarini.ea.2014} and \citet{Ghil.2019,Ghil2015}. 

\section{The Climate As a Nonequilibrium System}
Nonequilibrium statistical mechanics has made substantial progresses in recent decades \cite{G06,G14} and, as we will see below, the investigation of the climate system seems to be a perfect setting for applying its tools and for finding intellectual challenges for developing new general ideas. The nonequilibrium conditions of the climate system are primarily set by the inhomogeneous absorption of solar radiation \cite{Peixoto1992}. In addition, the climate system also receives a direct mechanical forcing from  solar and lunar tides. Such a forcing, while indeed important for some specific phenomena, is much less relevant than the radiative forcing coming from the Sun and can be altogether neglected in the discussion below. 

The absorption of solar radiation  preferentially takes place a) near the surface rather than in the deeper levels of the atmosphere and of the ocean; and b) at low rather than at high latitudes. An approximate steady state is reached in the climate system through a complex set of dynamical and thermodynamical processes that reduce the temperature gradients that would be established were only radiative processes involved \cite{Lucarini.ea.2014}. By and large, convective motions are key to reducing the inhomogeneity resulting from a), while large scale atmospheric and ocean heat transport is responsible for reducing the equator-to-pole temperature difference associated with b) above, thus providing a mechanism of global stability to the system. The hydrological cycle plays a fundamental role in terms of energetics of the climate system, mainly because of the large latent heat associated with water phase changes and of the possibly very large spatial scale associated with water transport processes \cite{Pauluis2,LucariniRagone,Trenberth2009,Lucarini2010b}.

It is possible to provide a succinct view of climate dynamics based on thermodynamical arguments: the climate can be seen as an imperfect engine able to transform available potential energy associated with temperature gradients into kinetic energy in the form of winds and oceanic currents through a variety of dynamical processes. The dominating processes are convective instability near the tropics, and baroclinic (and, to a much lesser extent, barotropic) instability in the mid-latitudes \cite{Peixoto1992,vallis_atmospheric_2006}. The kinetic energy  is continuously dissipated through various mechanism of friction, while available potential energy is dissipated through diffusive processes. This is the so-called Lorenz energy cycle \cite{Lor55,Lor67}, which can be augmented by defining the efficiency and the entropy production of the climate \cite{Kleidon05,Ambaum,Pauluis,Pauluis0,Lucarini:2009_PRE,Laliberte2015}. The Lorenz energy cycle angle on the dynamics of climate leads to clearly pointing out the separate roles played by the two main geophysical fluids. To a first approximation, the atmosphere is heated from below, and so is thermodynamically active \cite{Lor67,Tapio2006}, while the ocean is heated from above, which leads to imposing - by and large - a stable stratification. As a result, the ocean currents are mainly mechanically driven by surface winds, even if a non-trivial role is played also by localised density perturbations at the surface \cite{DG05,Kuhlbrodt2007,Cessi.2019}. As of today, at least in the author's opinion, despite the many merits of the description of the climate system as a thermal engine, a comprehensive and closed theory of climate dynamics able to explain coherently instabilities and stabilization mechanisms on the basis of the fundamental astronomical, physical, chemical, and geometrical parameters of the Earth system has not  yet been formulated.

\section{The Climate as a Multiscale System}
The nature of the external forcings, the inhomogeneity of the physical and chemical properties of the components of the climate system, as well as the great variety of dynamical processes occurring within each climatic component and of the coupling mechanisms between different components lead to the presence of non-trivial variability on a vast range of scales, covering over ten orders of magnitude in space - from Kolmogorov's dissipation scale to the Earth's radius - and even more than that in time - from microseconds to hundreds of millions of years \cite{Ghil2002}. Our knowledge of the system is extremely limited in terms of observational data: direct measurements of the climate system obtained with different and evolving technology, and thus having a moderate amount of synchronic and diacronic coherence, are available, to a first approximation, only in the industrial era \cite{Ghil.Mal.1991}, whereas for the more distant past one can only resort to indirect measurements in the form of proxy data \cite{Cronin2010}. The partial and inaccurate observational data are merged dynamically with a numerical model describing the evolution of the geophysical fluids through the process of data assimilation, whose aim is to provide the best time-dependent estimate of the state of the system given a set of available observations \cite{Ghil.Mal.1991,kalnay2003,Carrassi2018}.  

Additionally, it is unthinkable, given our current scientific understanding at large and our available or foreseen technological capabilities, to create a numerical model able to directly simulate the climate system in all details for a time frame covering all the relevant time scales. Furthermore, following the Poincar\'e parsimony principle, even if we had such a model, it would not serve the scope of advancing scientific knowledge, but would rather be a virtual reality emulator that would overwhelm a user by details to the point of obscuring the overall understanding of the problem; see a related discussion in, e.g., \citet{Held.gap.05}. As a result, each numerical model used for studying the climate system is formulated in such a way that only a (very limited) range of scales and processes are directly simulated, whereas the rest are either approximately parametrized and/or used to define suitable boundary conditions. Correspondingly, in order to study particular classes of phenomena, approximate evolution equations - which provide the basis for the numerical  modelling - are derived from the fundamental equations describing the dynamics of climate (basically, Navier-Stokes equations for multicomponent and multiphase thermodynamical fluids in a rotating frame of reference with a vast array of time-dependent forcings and non-trivial boundary conditions) in order to filter out certain physical processes that are heuristically assumed to play only a minor role at the temporal and spatial scales of interest \cite{vallis_atmospheric_2006,klein2010,Holton}. 

Therefore, the problem of constructing accurate and efficient reduced-order models (or, equivalently, of defining the coarse-grained dynamics) is an essential and fundamental aspect of studying the dynamics of climate, both theoretically and through simulations. Traditionally, parametrization schemes are formulated in such a way that one expresses the impact on the scales of interest of processes occurring within the unresolved scales via deterministic functions of the resolved variables. It has more recently become apparent, in the spirit of what is implied by the Mori-Zwanzig projection operator \cite{mori_transport_1965,zwanzig_memory_1961}, that, instead, parametrizations should involve  stochastic and non-markovian components \cite{palmer_stochastic_2009,Franzke.ea.2015,Berner2017}. Many strategies for constructing theoretically rigorous parametrizations have been devised, which can be broadly divided into top-down - see e.g., \citet{wouters_disentangling_2012,wouters_multi-level_2013,Vissio2018a,Wouters_2019a,Majda2001,Majda2013} - and data-driven approaches - see, e.g., \citet{wilks_effects_2005,KravtsovKondrashovGhil_JCL05,Kondrashov.Kravtsov.ea.2006,MSM2015}).

\section{Climate Variability and Climate Response}
One needs to remark that, additionally, the climate system is only in an approximate steady state, because the incoming radiation is subject to quasi-random  (e.g. sunspots, solar flares) as well as slow quasi-periodic modulations (e.g. Milankovitch cycles), and, on very long time scales, to changes in the intensity of the solar irradiance, resulting from the Sun's evolution. Additionally, the boundary conditions of the system change at a very slow pace in correspondence to geological processes, while the atmospheric composition is affected by volcanic eruptions, and, in more recent times, by Humanity itself, which acts as very rapid geological agent \cite{saltzman_dynamical}. Understanding climate change and its relationship to unperturbed natural climate variability is a grand challenge for contemporary science, with clear implications on:
\begin{itemize}
    \item better understanding and predicting how the ongoing anthropogenic climate change will manifest itself at different spatial and temporal scales, and how it will impact different subdomains of the climate system;
    \item gaining a more detailed knowledge on the co-evolution of the Earth’s climate and of life on Earth;
    \item better defining planetary habitability, i.e. the potential to develop and maintain environments hospitable to life (at least in the form we know or can envision) in a planet or in a satellite. 
\end{itemize}

When trying to relate forced  and free variability of a system, the default option is to try to use (one of the variants of) the fluctuation-dissipation theorem \cite{Kubo.1966,marconi2008}. Indeed, the fluctuation-dissipation theorem, which was originally formulated for systems that are near thermodynamic equilibrium, can be seen as a dictionary able to translate the statistics of free (thermal) fluctuations into a prediction of the response of the system to external perturbations. In the case of the climate system, this amounts to saying, \textit{grosso modo}, that the statistical properties of a different climate can be reconstructed by changing the statistical weight of the natural modes of variability of the reference climate. 

This viewpoint seems unable to account for the possibility of climatic surprises, i.e. the occurrence in the perturbed climate of events that were  absent in the reference case, as in the case of erratic variations in extreme events. This aspect hints at the need of adopting a slightly different approach when relating climate variability and climate change \cite{gritsun2017}. 

Indeed,  Ruelle's response theory \cite{ruelle_nonequilibrium_1998,ruelle_review_2009}  - see \citet{liverani2006} for a rigorous functional analysis viewpoint - indicates that for nonequilibrium systems obeying deterministic dynamics the fluctuation-dissipation theorem does not hold in its usual formulation. A nontrivial relationship between unforced fluctuations and response of the system to perturbations can be recovered by using the formalism of the transfer operator and studying the properties of the so-called Ruelle-Pollicott poles \cite{Pollicott1985,Ruelle1986} of the unpertubed system \cite{Lucarini2018JSP,Chekroun2014}\footnote{A rigorous formulation of response theory in the context of stochastic dynamics has been proposed by \citet{Hairer2010}. Recently, \citet{Wormell2019} have clarified the link between the deterministic and the stochastic viewpoints. Note that in the case of stochastic systems (a general form of) the fluctuation-dissipation theorem is valid \cite{marconi2008}. See \citet{Gottwald2020} for a recent special issue on linear response theory and its applications.}.

An extremely fascinating aspect of climate variability is associated with the occurrence of extreme events, such as heat waves, cold spells, droughts, floods, wind storms, and many others. Extreme meteo-climatic events can be wildly different in terms of spatial and temporal scales of interest (e.g. droughts are typically associated with much longer time scales and much large spatial scales than floods) because of the variety of physical processes responsible for them. The special importance given to the study of extremes in climate comes essentially from their relevance in terms of impacts - while, by definition, rare in terms of occurrence, they are disproportionally responsible for damage inflicted to society and ecosystems \cite{IPCC12}. Extreme value theory \cite{Coles.2001} allows for a detailed description of extreme meteo-climatic events \cite{katz2005statistics,ghil2011}. Recently, it has been shown that the investigation of extremes allows for understanding the dynamical properties of the system generating them \cite{LKFW14,lucarini2016extremes}. These results are finding applications for providing a new viewpoint for the investigation of atmospheric  predictability \cite{Faranda2017}. The application of large deviation theory in the study of the climate is rather recent, and it has shown great potential in describing the properties of persistent extreme events like heat waves \cite{Galfi2019}. At a more abstract level, large deviation theory-based tools have been instrumental in nudging climate model simulations towards very rarely explored regions of the phase space, thus enhancing tremendously the possibility of studying mechanisms behind extreme events \cite{Ragone2017}. 

Finally, we have to keep in mind that we should abandon the hypothesis of considering the climate as an autonomous system, and consider instead the impact of stochastically and deterministically varying paramemeters \cite{Ghil2015}. A suitable mathematical setting for studying the statistical properties of the climate system is given by  pullback (rather than regular) attractors \cite{GCS08,Chekroun2011,CLR13}, which are the support of a time-dependent measure. While it is possible to provide precise mathematical definitions for the pullback attractor, the construction of the corresponding measure in a given numerical model and its use for computing the time-dependent values of observables of interest is very challenging at practical level, because one needs many ensemble members for approximating the actual measure with the empirical one. Under suitable hypotheses of structural stability - namely the chaotic hypothesis \cite{gallavotti_dynamical_1995} - Ruelle's response theory appears, despite the difficulties in constructing the response operators \cite{abramov2007}, as an efficient and  flexible tool for calculating climate response to weak and moderate forcings, greatly generalising classical concepts like equilibrium climate sensitivity (long term change in the globally averaged surface temperature as a result of doubling in the CO$_2$ concentration \cite{IPCC13,vonderHeydt2016}), for explaining \cite{Nijsse2018} the theory of emergent constraints \cite{Collins2012} and, more generally, for reconstructing the properties of the pullback attractor from a suitably defined reference background state \cite{Lucarini2017,Lembo2020}; see also an interesting application in \cite{Aengenheyster2018}. Changes in the statistics of extreme events can also be predicted using Ruelle's response theory \cite{LKFW14}.

\section{The Climate Crisis: Non-smooth Climate Response and Critical Transitions}

The previous discussion, as well the usual narrative about climate change, often gives the impression that climate change manifests itself essentially through gradual modulations. Time goes by, the CO$_2$ concentration increases, and the temperature goes up. In this regard, the expression  \textit{climate crisis} has recently become prominent in the public discourse and some prominent media outlets do not use anymore the expression \textit{climate change}  (see, e.g., the statement made by editorial board of \textit{The Guardian} at \texttt{https://tinyurl.com/yyu54ae5}). On the one hand, \textit{climate crisis}   evokes a strong emotional response and conveys a sense of urgency; on the other hand, it is, indeed, a very appropriate technical term, because one of the most pressing challenges in climate science is achieving a much deeper understanding of its critical transitions \cite{scheffer2009critical,Kuehn2011}, which are usually referred to as \textit{tipping points} in the Earth system science jargon \cite{Boers2017,Lenton.tip.08,Ashwin2012}. These can be seen, by and large, as nonequilibrium phase transitions leading to drastic and possibly catastrophic changes in the climate. When a system nears a critical transition, its properties have a rough  dependence on its parameters, because response to perturbations is greatly enhanced \cite{Chekroun2014}. Conversely, the radius of expansion of response theory becomes very small \cite{Lucarini2016} and the decay of correlation for physically meaningful observables slows down, as a result of a vanishing spectral gap associated to the subdominant Ruelle-Pollicott pole(s) \cite{Tantet2018}. 

Critical transitions in the climate system are especially relevant because they often accompany the property of multistability. If one considers the case of deterministic dynamics, in a certain range of values of the parameters of the system, there are two or more competing steady states that can be reached by the system. Important climatic subsystems such as the Atlantic Meridional Overturning Circulation \cite{Rahmstorf2005} and the Amazon ecosystem \cite{Wuyts2017} are considered to be bistable. And the climate as a whole is indeed multistable, as the current astrophysical and astronomical conditions support at least two possible climates - the one we live in, and the ice-covered one, often referred to as snowball state \cite{Budyko,Sellers,Ghil1976,HoffmanSchrag,Boschi2013}. 

The competing steady states are associated with attractors that are the asymptotic sets of orbits starting inside their corresponding basins. On the boundary of such basins we have invariant sets, the Melancholia states, that attract initial conditions on the basin boundary. These states can be constructed using the edge tracking algorithm originally devised for studying turbulent fluids \cite{Skufca2006}. Similarly to the case of simple gradient systems, such unstable saddles define the global stability properties of the system \cite{Grebogi1983,LT:2011,LucariniBodai2017}. The Melancholia states are the gateways of noise-induced transitions, no matter the noise law one selects. Large deviation theory \cite{T09} provides us with tools to compute the invariant measure of the system and the statistics of transition times between different metastable states \cite{LucariniBodai2019PRL,LucariniBodai2019arxiv}.

\section{This Special Issue}
We present below a brief narrative of this special issue, and propose a rough thematic grouping of the included contributions, which cover most of the topics mentioned in the previous sections as well as exploring further exciting research directions. 
\begin{itemize}
    \item \citet{WeissJSP} provide a conceptually elegant characterisation of the nonequilibrium properties of the climate system by computing persistent probability currents, which cannot be found in equilibrium systems. The corresponding current loops are used to characterise key climatic features like the El-Ni{\~n}o Southern Oscillation (ENSO) and the Madden--Julian Oscillation  and define a new indicator in the form of a probability angular momentum. Instead of focusing on steady state properties, \citet{GottwaldJSP} propose a data-driven method based on the Koopman operator to detect eventual regime changes and fast transient dynamics in time series, and provide convincing applications in a chaotic partial differential equation and in atmospheric data of the Southern and of the Northern Hemisphere. Equilibrium statistical mechanics is instead the framework of the contribution by \citet{ContiJSP}, who investigate the so-called generalized Euler equations, including those describing surface quasi-geostrophic dynamics, which is especially relevant for small scale oceanic features associated with horizontal gradients of buoyancy. They propose a generalised selective decay principle able to explain the equilibrium state of the flow. Such a principle imposes that the solutions of these equations approach the states that minimise the generalized potential enstrophy compatibly with the  value of the generalized energy.

    \item The investigations of the multiscale nature of climate is a \textit{fil rouge} of this special issue. \citet{ChekrounJSPVariantional} proposes an innovative framework for constructing nonlinear parameterizations of unresolved scales of motions using a variational approach and presents applications in the context of the primitive equations describing the dynamics of the atmosphere and the Rayleigh--B{\'e}nard convection. \citet{TondeurJSP}, conversely, explores new aspects of coupled data assimilation schemes able to merge observational and model generated data pertaining to from both the atmosphere and the ocean. A model reduction technique based on a stochastic variational approach for geophysical fluid dynamics is used for developing a new ensemble-based data assimilation methodology for high-dimensional fluid dynamics models in \citet{CotterJSP}. \citet{EichingerJSP}, instead, address the impact of adding noise in the form of additive fractional Brownian motion on fast-slow systems and investigates the problem of estimating how likely is for the trajectories to stay nearby the slow manifold of the deterministic system.

    \item A new paradigm for climate science is suggested in three closely linked contributions. \citet{AlonsoJSP} propose in the case of two-dimensional Euler-Boussinesq equations a closed theory of weather and climate - intended as statistics of fluctuations and expectation value of the quantities of interest, respectively. This is achieved by taking into account the corresponding \textit{Lagrangian averaged stochastic advection by Lie transport} equations, which are nonlinear and non-local, in both physical and probability space \cite{DrivasJSP}. This formalism is further used to explore the properties of a stochastically perturbed low-dimensional chaotic dynamical system. \citet{GeurtsJSP} show how to construct an effective dynamics for the expectation values of the solutions, which can be mapped to the original deterministic system by considering a renormalised dissipation.
    
\item A three-part contribution tries to address the overall problem of inferring information on the natural variability of a system by studying the Ruelle-Pollicott resonances extracted from time series of partial observations. The overall theoretical framework is presented in \citet{ChekrounJSPI}, while in \citet{TantetJSPII} it is shown how to investigate Hopf bifurcations for a stochastic system and  extract from data a clear fingerprint of  nonlinear oscillations taking place within a stochastic background. Furthermore,  the theoretical findings are exploited for studying the oscillations in the Cane-Zebiak stochastic ENSO model \cite{TantetJSPIII}.

\item The concept of natural variability requires using a different framework when considering nonautonomous systems. As discussed above, the characterisation of the properties of the pullback attractor requires considering ensemble simulations, and it is in general not clear how large the ensemble should be in order to be able to make meaningful statements in the statistical properties of the system. This issue is addressed by \citet{PieriniJSP} using a simplified model of the ocean. Time-dependent correlations between the parallel climate realizations defining the pullback attractor are used in \citet{TelJSP} to provide an alternative definition for teleconnections, i.e. correlation between anomalies of climatic fields in regions that are geographically very far away.  

\item The problem of studying how the climate system responds to forcings is investigated in different directions.  \citet{AshwinJSP} contrast the usual definition of equilibrium climate sensitivity, which relies, as discussed above, on taking a linear approximation to the climate response, with a fully nonlinear version, and aim at studying the behaviour of such quantities in the presence of tipping points in simple climate models. Climate response is  investigated on paleoclimatic data by \citet{AhnJSP} with the goal of understanding whether it is possible to define causal links  between different proxy records as a result of the presence of cross-correlations, and discuss the need for conjecturing the presence of a separate forcing responsible for the observed signals. The problem of predicting climate response to forcings motivates the study by \citet{SantosJSP}, who provide general formulas for computing linear and nonlinear response to fairly general forcings in the context of finite Markov chains and use them, after performing a discretisation of the phase space, to study the sensitivity of a deterministic and a stochastic dynamical system. Furthermore, \citet{MarangioJSP} study the linear response of a stochastically forced Arnold circle map with respect to the frequency of the driving frequency. This map is taken as a climate toy model, and the goal of the study is, specifically, to gain insights into the ENSO phenomenon. Finally, the problem of exploring the multistability properties of the climate system acts as primary motivation for the study by \citet{BodaiJSP}, who proposes a new algorithm for evaluating efficiently in a stochastically perturbed system the quasi-potential barrier that confines a given metastable state and tests the approach in a classical energy balance climate model.

\item The study of extreme events is a key aspect of several contributions to this special issue. \citet{RagoneJSP} discuss critically how concepts and algorithms informed by large deviation theory can be applied to study persistent extreme events in climate and present an application focusing on persistent and spatially extended heat waves performed using a comprehensive  climate models. A different angle on persistent extremes is given in the study by \citet{VaientiJSP}, who provide a very informative review of the extremal index. The extremal index is a quantity one defines in the context of extreme value theory that can be used to quantify the presence of clusters of extremes events, and can, in the context of dynamical systems, provide information on local  and global properties of the attractors. \citet{VaientiJSP} study both the case of deterministic and of stochastic dynamical systems, and present also an example of analysis of actual atmospheric data. Extreme value theory is used by \citet{PonsJSP} to propose a way to address well-known problem of the curse of dimensionality in estimating the Hausdorff dimension of the attractor from time series. The methodology is applied to study recurrences in synthetically generated as well as real-life financial and climate data and define the \textit{degree of non-randomness} of the system. Finally, motivated by the interest in extreme events near coastal continental shelves for water waves, \citet{MajdaJSP} investigate the change in the statistical features found when looking at water surface waves going across an abrupt depth change. The flow is modelled using a truncated version of the Korteweg–de Vries equation and the transition between the statistical regimes of the incoming and of outgoing field is studied by constructing mixed Gibbs measures with microcanonical and canonical components. 

\end{itemize}

\section*{Acknowledgments}
VL wishes to thank all the authors of the special issue for their fascinating contributions; Elena Baroncini, Alessandro Giuliani, and Aldo Rampioni for their support and patience in the preparation of this special issue; Giovanni Gallavotti, Michael Ghil, Joel Lebowitz, and David Ruelle for amazing intellectual inspiration and personal exchanges.  VL  acknowledges the support received from the EU Horizon2020 project TiPES (grant No. 820970). VL takes the liberty of dedicating this special issue to Vito Volterra (1860-1940), who had the rare gift of matching scientific genius with moral strength and steadfastness against fascism. As a trailblazing interdisciplinary scientist who pioneered  mathematical modelling for environmental sciences, he would have probably enjoyed the current developments of the mathematics and physics of climate.  

\end{document}